\documentstyle[12pt]{article}

\title{{\normalsize{{\hskip 8.5cm} BIHEP-TH-96-19}}\\
      QCD Sum Rules for The Double Ratio 
$(f_{B_{s}}/f_{B_{d}})/(f_{D_{s}}/f_{D_{d}})$ in HQET}
\author{Tao Huang$^{1,2}$, Zuo-Hong Li$^{2}$ and Chuan-Wang Luo$^{2}$\\
       {\small $^{1}$CCAST(World Laboratory)P.O.Box 8730,
        Beijing 100080, P. R. China,}\\
        {\small $^{2}$Institute of High Energy Physics, P.O.Box 918(4),
        Beijing 100039, P. R. China.\thanks{\small mailing address}}}
\date{}

\begin{document}
\maketitle
\vspace{2cm}
\begin{abstract}
The double ratio $(f_{B_{s}}/f_{B_{d}})/(f_{D_{s}}/f_{D_{d}})$ 
is calculated by QCD sum rules
in heavy quark effective theory (HQET),
both numerically and analytically. Our expression for the double ratio shows 
explicitly the dependence on the light quark masses, the heavy quark
symmetry breaking and the vacuum condensates. The numerical result favors
the double ratio to be a little greater than 1.
\end{abstract}


\newpage
\hspace{0.35cm}Precision determination of the double ratio 
$(f_{B_{s}}/f_{B_{d}})/(f_{D_{s}}/f_{D_{d}})$ is of important interest
in understanding the $B-\bar{B}$ mixing phenomena.
Both lattice and
QCD sum rule calculations\cite{1,2} have given that $f_{B_{s}}/f_{B_{d}}$
and $f_{D_{s}}/f_{D_{d}}$ are close to 1. 
Grinstein\cite{3}, based on the phenomenological Lagrangian incorporating 
heavy quark and chiral symmetries,
has found the double ratio $(f_{B_{s}}/f_{B_{d}})/(f_{D_{d}}/f_{D_{d}})=0.967$.
Oakes\cite{4} has also done similar work 
by making use of the Wigner-Eackart theorem concluding that 
$f_{B_{s}}/f_{B_{d}}=0.989$, $f_{D_{s}}/f_{D_{d}}=0.985$ and 
$(f_{B_{s}}/f_{B_{d}})/(f_{D_{d}}/f_{D_{d}})=1.004$. Also,
It is noted that there is a difference in the sign of correction term
away from 1 between Grinstein's and Oakes's expressions for the double ratio. 
Combination of QCD sum rule method with HQET, as a powerful approach
to nonperturbative dynamics in the weak decays of heavy hadrons,
provides us with a possibility to discuss
the double ratio from QCD theory.
In the previous paper\cite{5}, we have estimated the decay constants
of heavy mesons and their SU(3) breaking  
effects by using QCD sum rule in HQET, obtaining
$f_{B_{s}}/f_{B_{d}} = 
1.17\pm0.03$ and
$f_{D_{s}}/f_{D_{d}} = 1.13\pm0.03$.
Therefore, the double ratio can be extracted from the above results as
$(f_{B_{s}}/f_{B_{d}})/(f_{D_{s}}/f_{D_{d}}) = 1.035 \pm 0.06$.
Obviously, the indirect calculation brings about a large error. 
To get a more precise result,
in the present work
we try to discuss it directly within the QCD sum rule framework of HQET. 
It should be 
emphasized that in our approach the SU(3) symmetry breaking effect  
in the double ratio, as has been shown in Ref.\cite{5}, depends on
not only the current quark masses, but also the condensates
$<0|\bar{q}q|0>$ and $<0|\bar{q}\sigma Gq|0>$. This is different from
the investigations of Grinstein and Oakes, where the SU(3) symmetry
breaking is determined only by the current quark masses.

\hspace{0.35cm}let's begin with a brief review on 
Grinstein's
and Oakes's works. Decay constants of
heavy mesons $M$ are defined as 
\begin{equation} 
<0|\overline{q}\gamma_{\mu}\gamma_{5}Q|M>=if_{M}P_{\mu} .
\end{equation}
According to Grinstein\cite{3}, the ratio of decay constant, for instance,
$f_{B{s}}/f_{B{d}}$ 
, can be expanded
in a quantity $m_{s}/\Lambda$($\Lambda$ is $QCD$ scale)
if only $SU(3)$ breaking is taken into account, 
$f_{B_{s}}/f_{B_{d}}=1+\frac{m_{s}}{\Lambda}$.                       
Heavy quark symmetry breaking modifies this expression as
$f_{B_{s}}/f_{B_{d}}=1 + \frac{m_{s}}{\Lambda}(a_{0} + a_{1}\frac{\Lambda}
{m_{q}}+ \cdots)$. 
The double ratio incorporating SU(3)
and heavy quark symmetry breaking effects may read approximately
\begin{equation}
(f_{B_{s}}/f_{B_{d}})/(f_{D_{s}}/f_{D_{d}}) 
= 1 + a_{1}(m_{s}/m_{b} - m_{s}/m_{c}).
\end{equation}
Setting $a_{1}=1$ for the coefficient of the correction term and   
$m_{s}=150$MeV, $m_{c}=1.5$GeV and $m_{b}=4.5$GeV for the current quark 
masses, the double ratio is approximately $0.93$, which is smaller than 1.
To better estimate the double ratio, Grinstein\cite{3}, making use of 
the phenomenological Lagrangian for heavy mesons
and light pseudoscalars incorporating heavy quark and chiral symmetries,
found it to be $0.967$.
In Ref.\cite{4}, Oakes, utilizing the relation $\partial^{\mu}A_{\mu} = 
i(m_{Q} + m_{q})\bar{q}\gamma_{5}Q$, obtained exactly

\begin{equation}
f_{B_{s}}/f_{B_{d}} = \bigg(\frac{M_{B_{d}}}{M_{B_{s}}}\bigg)^{2}
\bigg(\frac{m_{b} + m_{s}}{m_{b} + m_{d}}\bigg)\frac{<0|
\overline{s}\gamma_{5}b|B_{s}>}{<0|\overline{d}\gamma_{5}b|B_{d}>} 
\end{equation} 
and
\begin{equation}
f_{D_{s}}/f_{D_{d}} = \bigg(\frac{M_{D_{d}}}{M_{D_{s}}}\bigg)^{2}
\bigg(\frac{m_{c} + m_{s}}{m_{c} + m_{d}}\bigg)\frac{<0|
\overline{s}\gamma_{5}c|D_{s}>}{<0|\overline{d}\gamma_{5}c|D_{d}>}.
\end{equation} 

\noindent Thus, the ratios of decay constants can be 
precisely estimated with the help of
the Wigner-Eackart theorem which allows 
the approximation relation $<0|\bar{s}\gamma_{5}Q|M_{Qs}> \approx
<0|\overline{d}\gamma_{5}Q|M_{Qd}>$ to hold, to very high accuracy,

\begin{equation}
f_{B_{s}}/f_{B_{d}} = \bigg(\frac{M_{B_{d}}}{M_{B_{s}}}\bigg)^{2}
\bigg(\frac{m_{b} + m_{s}}{m_{b} + m_{d}}\bigg)
\end{equation}
and
\begin{equation}
f_{D_{s}}/f_{D_{d}} = \bigg(\frac{M_{D_{d}}}{M_{D_{s}}}\bigg)^{2}
\bigg(\frac{m_{c} + m_{s}}{m_{c} + m_{d}}\bigg).
\end{equation}
From Eq.(5) and Eq.(6), the double ratio
$(f_{B_{s}}/f_{B_{d}})/(f_{D_{s}}/f_{D_{d}}) = 1.004$,
and an approximate expression 
can be found
\begin{equation}
(f_{B_{s}}/f_{B_{d}})/(f_{D_{s}}/f_{D_{d}}) \simeq 1 - 2\bigg(\frac{M_{B_{s}}
- M_{B_{d}}}{M_{B_{d}}} - \frac{M_{D_{s}}
- M_{D_{d}}}{M_{D_{d}}}\bigg) 
+ \frac{m_{s}
- m_{d}}{m_{b}} - \frac{m_{s}
- m_{d}}{m_{c}}. 
\end{equation}                
To the extent that $M_{B_{s}} - M_{B_{d}}\simeq M_{D_{s}} - M_{D_{d}} 
\simeq m_{s} - m_{d}\simeq m_{s}$, $M_{B}\simeq m_{b}$ and
$M_{D}\simeq m_{c}$, Eq.(7) is simplified as following
\begin{equation}                                                         
(f_{B_{s}}/f_{B{_d}})/(f_{D_{s}}/f_{D_{d}}) \simeq 1 + a\bigg(\frac{m_{s}}
{m_{b}} - \frac{m_{s}}{m_{c}}\bigg),
\end{equation}
which shows that the double ratio is, indeed, unity up to a correction
of order $a(m_{s}/m_{b} - m_{s}/m_{c})$, albeit with coefficient $a$
of -1 rather than 1.  

Now, we turn to calculation of an analytic form of the double ratio
by means of QCD sum rules in HQET. The heavy quark effective lagrangian
to next-leading order in $1/m_{Q}$ is given by\cite{6}
\begin{equation}
L= \overline{h_{v}}iv\cdot Dh_{v} + \frac{L_{K}}{2m_{Q}}
  + \frac{C_{mag}(\mu)L_{s}}{2m_{Q}},
\end{equation}
where $L_{K}=\overline{h_{v}}(iD)^{2}h_{v}$ 
is the kinetic
energy operator of heavy quark , $L_{s}= \frac{1}{2}\overline{h_{v}}
g_{s}\sigma_{\mu\nu}G^{\mu\nu}h_{v}$ is the chromomagnetic interaction operator
and $C_{mag}(\mu)= 
[\frac{\alpha_{s}(m_{Q})}{\alpha_{s}(\mu)}]^
{3/(11-2n_{f}/3)}[1+\frac{13}{6}\frac{\alpha_{s}}{\pi}]$ 
is the Wilson coefficient.
To calculate decay constants 
of heavy mesons
by HQET approach, matching current operators in fully theory
onto ones in HQET is necessary. Let's calculate the axial current
matrix element between the vacuum and a heavy meson state.  
A match\cite{5,7} of the axial current operator in fully theory onto the local
operators of the effective theory gives the expression
for the decay constant at order $1/m_{Q}$ in the heavy quark expansion in
the case of light quark mass $m_{q}\not=0$ 
\begin{equation}
f_{M}\sqrt{m_{M}} = \hat{F}(m_{Q})[1 + d_{M}\frac{\alpha_{s}(m_{Q})}
{6\pi}]\{1 + \frac{\hat{G}_{k}(m_{G})}{m_{Q}} +
\frac{2d_{M}}{m_{Q}}[\hat{G}_{\Sigma}(m_{Q}) - \frac{\bar{\Lambda}}
{12}]\},
\end{equation}
with $d_{M}=3$ for pseudoscalar mesons and $-1$ for vector mesons.
$\hat{F}(m_{Q})$,$\hat{G}_{K}(m_{Q})$ and $\hat{G}_{\Sigma}(m_{Q})$ 
are some renormalization group invariant quantities\cite{5}:
\begin{equation}
\hat{F}(m_{Q}) = [C_{1}(\mu)+\frac{C_{2}(\mu)}{4}]F(\mu),
\end{equation}                                                    
\begin{equation}
\hat{G}_{K}(m_{Q}) = G_{K}(\mu) - \frac{\overline\Lambda}{6}b(\mu)
+ \frac{m_{q}}{6}a(\mu)
\end{equation}
and
\begin{equation}
\hat{G}_{\Sigma}(m_{Q}) = G_{mag}(\mu)G_{\Sigma}(\mu) - 
\frac{\overline{\Lambda}}{12}[B(\mu)-1] 
+ \frac{m_{q}}{12}A(\mu).
\end{equation}
In the above equations,
the short$-$distance coefficients $C_{1}(\mu)$, $C_{2}(\mu)$, $a(\mu)$, 
$A(\mu)$,
$b(\mu)$ and $B(\mu)$ have been obtained in Ref.\cite{5}. $F(\mu)$, 
$G_{K}(\mu)$, 
$G_{\Sigma}(\mu)$ and $\bar{\Lambda}$ are some long$-$distance parameters
\cite{5,7}:   
$F(\mu)$ is used to parametrize the leading order
matrix element  
in $1/m_{Q}$ expansion, $G_{K}(\mu)$ and $G_{\Sigma}(\mu)$ are two 
additional universal parameters, which are respectively
introduced to parametrize the $1/m_{Q}$ corrections to the hadronic
wave function by insertions of the subleading operators $L_{K}$ and 
$L_{\Sigma}$ in the effective lagrangian into the matrix element
of the leading order current, and $\bar{\Lambda}$ expresses mass 
difference of the heavy meson and the heavy quark inside it.
It is interesting to note there is an explicit dependence of these 
low energy parameters on not only the light quark masses
but also the condensate parameters in our approach, as shown in
Ref.\cite{5}. Considering these effects, we can obtain easily the expressions 
for the ratios of decay constants 
\begin{equation}
f_{B_{s}}/f_{B_{d}} = \frac{F_{s}(\mu)\sqrt{m_{B_{d}}}}  
{F_{d}(\mu)\sqrt{m_{B_{s}}}}
\frac{1 + \frac{C_{b,s}(\mu)}{m_{b}}}{1 + \frac{C_{b,d}(\mu)}{m_{b}}} 
\end{equation}
and
\begin{equation}
f_{D_{s}}/f_{D_{d}} = \frac{F_{s}(\mu)\sqrt{m_{D_{d}}}}       
{F_{d}(\mu)\sqrt{m_{D_{s}}}}
\frac{1 + \frac{C_{c,s}(\mu)}{m_{c}}}{1 + \frac{C_{c,d}(\mu)}{m_{c}}} 
\end{equation}
with $C_{Q,q}(\mu) = G_{K}^{q}(\mu) + 6C_{mag}^{Q}
(\mu)G_{\Sigma}^{q}(\mu)
+ \frac{a(\mu,m_{Q})}{6}m_{q} - \frac{b(\mu,m_{Q})}{6}\bar{\Lambda}_{q}$. 
Following the standard method for QCD sum rule calculation\cite{8},
one can get the sum rules for those long$-$distance coefficients\cite{5}:
\begin{equation}
\bar{\Lambda}_{q} = -\frac{1}{2}\frac{\partial}{\partial T^{-1}}lnK ,
\end{equation}
\begin{equation}
 F_{q}(\mu)e^{-2\bar{\Lambda}_q/T} = K ,
\end{equation}
\begin{equation}
G_{K}^{q}(\mu) = \frac{1}{2}\frac{d}{dT}\bigg(\frac{TJ_{K}}{K}\bigg) 
\end{equation} 
and
\begin{equation}
G_{\Sigma}^{q}(\mu) = \frac{1}{4}\frac{d}{dT}\bigg(\frac{TJ_{K}}{K}\bigg).
\end{equation}
Here, we have not given the final expressions\cite{5} for $K$, $J_{K}$ 
and $J_{\Sigma}$,
which are dependent on the Borel parameter $T$, threshold 
$\omega^{c}$ of the continuum states and light flavors.
In the light of Eq.(14) and Eq.(15) and by a simple manipulation, 
we have the double ratio
\begin{eqnarray}
(f_{B_{s}}/f_{B_{d}})(f_{D_{s}}/f_{D_{d}}) = 1 -&G(\mu,m_{b})
\frac{m_{s}-m_{d}}{m_{b}} + G(\mu,m_{c})
\frac{m_{s}-m_{d}}{m_{c}}\nonumber\\ + &\Delta G(\mu,m_{b})\frac{1}{m_{b}}
- \Delta G(\mu,m_{c})\frac{1}{m_{c}} 
\end{eqnarray}
to next-leading order in $1/m_{Q}$, where $G(\mu,m_Q)$ is defined as
\begin{equation}
G(\mu, m_{Q})= \frac{1}{2}[B(\mu,m_{Q})- A(\mu,m_{Q})]         
             + \frac{1}{6}[b(\mu,m_{Q})- a(\mu,m_{Q})]    
\end{equation}
and
$\Delta G(\mu, m_{Q})$ is SU(3) symmetry breaking parameter
\begin{equation}
\Delta G(\mu, m_{Q})= \Delta G_{K}(\mu) + 6C_{mag}(\mu,m_{Q})         
                    \Delta G_{\Sigma}(\mu)
\end{equation}
with
\begin{equation}
\Delta G_{K}(\mu) = G_{K}^{s}(\mu)- G_{K}^{d}(\mu)  
\end{equation}
and
\begin{equation}
\Delta G_{\Sigma}(\mu) = G_{\Sigma}^{s}(\mu)- G_{\Sigma}^{d}(\mu).
\end{equation}
In deriving Eq.(20),
the approximate relations $M_{B_{s}}- M_{B_{d}}\simeq     
M_{D_{s}} - M_{D_{d}}\simeq m_{s}-m_{d}$ and $\bar{\Lambda}_s - \bar{\Lambda}_d
\simeq m_{s}-m_{d}$ have been used. 

The Wilson coefficients $G(\mu,m_{Q})$ are easily evaluated
via taking $\mu=1GeV$, $m_b=4.5\sim5.0$ GeV and $m_c=1.2\sim1.5$ GeV. 
Now we are in a position to compute
two low energy parameters $\Delta G_{K}(\mu)$ and $\Delta G_{\Sigma}(\mu)$, 
by resorting to the QCD sum rules in HQET.  
Before numerical manipulation,
we would like to deal simply with these two quantities.
If we assume both $G_{K}^{q}(\mu)$ and $G_{\Sigma}^{q}(\mu)$ to 
be continuous functions in the light quark mass $m_{q}$,        
$\Delta G_{K}(\mu)$ and $\Delta G_{\Sigma}(\mu)$
can be expanded in terms of $(m_{s}-m_{d})$ respectively
\begin{equation}
\Delta G_{K}(\mu) = f_{K}(m_{s}-m_{d}) +\cdots 
\end{equation}
and                                                                 
\begin{equation}
\Delta G_{\Sigma}(\mu)= f_{\Sigma}(m_{s}-m_{d}) +\cdots ,
\end{equation}
where $f_{K}$=$\frac{dG_{K}}{dm_{q}}|_{m_{q}=m_{d}}$,              
$f_{\Sigma}$ =$\frac{dG_{\Sigma}}{dm_{q}}|_{m_{q}=m_{d}}$ and        
$``\cdots "$ denotes the higher power terms of $(m_{s}-m_{d})$. 
If we neglect contributions from the higher power terms, Eq.(20)
becomes
\begin{equation}
(f_{B_{s}}/f_{B_{d}})/(f_{D_{s}}/f_{D_{d}}) = 1 + a_{1}
\bigg(\frac{m_{s}-m_{d}}{m_{b}}\bigg)\nonumber\\ 
 - a_{2}
\bigg(\frac{m_{s}-m_{d}}{m_{c}}\bigg) 
\end{equation}
with $a_{1} = f_{K}+6C_{mag}(\mu,m_{b})f_{\Sigma}-G(\mu,m_{b})$
and
$a_{2} = f_{K}+6C_{mag}(\mu,m_{c})f_{\Sigma}-G(\mu,m_{c})$.
Making an approximation $m_{s}-m_{d}\approx m_{s}$, Eq.(27) can be simplified 
as
\begin{equation}
(f_{B_{s}}/f_{B_{d}})/(f_{D_{s}}/f_{D_{d}}) = 1 + a_{1}
\frac{m_{s}}{m_{b}}\nonumber\\ 
 - a_{2}
\frac{m_{s}}{m_{c}} . 
\end{equation}
The small current quark mass $m_d$ allows us to evaluate precisely
$f_{K}$ and $f_{\Sigma}$
at $m_{q}\approx0$. 
Along the same line as in Ref.\cite{5},
it is straightforward to get the sum rules for $f_{K}$ and $f_{\Sigma}$
\begin{equation}
f_{K} = \frac{1}{2}\frac{d^{2}}{dm_{q}dT}\bigg(\frac{TJ_{K}}{K}\bigg),
\end{equation}
and
\begin{equation}
f_{\Sigma} = \frac{1}{4}\frac{d^{2}}{dm_{q}dT}
\bigg(\frac{TJ_{\Sigma}}{K}\bigg).
\end{equation}
For simplicity, we do not give the final expressions 
for these two coefficients. To calculate these two coefficients at $m_{q}
\approx0$, the formulas\footnote{$\frac{d}{da}\int_{\psi(a)}   
^{\varphi(a)}f(x,a)dx = f[\varphi(a),a]\frac{d\varphi(a)}{da}-           
f[\psi(a),a]\frac{d\psi(a)}{da} +
\int_{\psi(a)}^{\varphi(a)}\frac{df(x,a)}{da}dx$}
for differential calculus can be used. However, an exact relation
between the threshold $\omega^c$ and the light quark mass $m_q$ is unknown. 
A resonable and simple
assumption for this relation, according to the definition of the $\omega$ 
in HQET, 
is $\omega^c(m_q)=\omega^c(0) + Am_q$     
with a coefficient $A$ varying between 0.0 and 2.0. In fact, this linear
relation has successfully be applied to QCD sum rule calculations in
Ref.\cite{5}.
Using the same energy scale $\mu=1$GeV and condensate 
parameters as in Ref.\cite{5},
we can carry out the sum rule calculations for $f_{K}$
and $f_{\Sigma}$, and further obtain our desired results for $a_1$ and $a_2$. 
In the manipulation process, we find a stable window 
$T=0.7\sim1.0$ GeV to exist for the threshold $\omega^c=2.0\pm0.3$ GeV.
Our numerical result shows 
that $a_1$ and $a_2$ are insensitive to variation of $A$ from
0.0 to 2.0, because of large cancellation effects between 
the first two terms in the expressions for $a_1$ and $a_2$.
For instance, when taking A = 0.0, 0.5, 1.0, 1.5 and 2.0, we 
have $a_{1} = - 1.188, -1.201, -1.208, -1.213$ and $-1.215$, 
and $a_2 = -0.905, -0.916, -0.921, -0.924$ and $-0.925$, respectively
(an error of $4\%$ in each exists due to the 
uncertainty of the comtinuum threshold $\omega^c$). 
In addition, it is worthwhile to emphasize that the uncertainties of 
the quark masses produce only the corrections of about $2\%$ to $a_{1}$ and 
$a_{2}$. 
The two coefficients $a_1$ and $a_2$ in case of A=1, 
as the functions of the Borel parameter T, 
are plotted in Fig.1. 

To summarize, in this letter, we discuss the double ratio$(f_{B_{s}}/f_{B_{d}})
/(f_{D_{s}}/f_{D_{d}})$ by applying QCD sum rules in HQET.
We find that the correction term $a_{1}(m_{s}/m_{b}-m_{s}/m_{c})$ 
for the double ratio in Ref.\cite{3} and Ref.\cite{4}
is modified
to $a_{1}(m_{s}/m_{b}) - a_{2}(m_{s}/m_{c})$ slightly, with $a_1\approx -1.2$
and $a_2\approx -0.9$. Our expression for the double ratio differs
from those of Grinstein and Oakes in two aspects:
(i) the double ratio in our approach 
is determined not only by quark mass terms but also by vacuum condensates,
while in Ref.\cite{3} and Ref.\cite{4} 
the SU(3) breaking effect in the double ratio results only from
the light quark masses; (ii) as a result in HQET,
$a_1$ and $a_2$ depend weakly on the heavy quark masses $m_b$ and $m_c$, 
respectively, via $\alpha_s(m_b)$ and
$\alpha_s(m_c)$. This leads to the small difference between $a_1$ and $a_2$.
Considering uncertainties of the heavy meson
and the quark masses, we can conclude that the double ratio
is a little greater than 1. This is also consistent with Oakes's prediction.  
Our conclusion will be beneficial to precision determination of
$f_{B_{s}}/f_{B_{d}}$ from the data on $f_{D_{s}}/f_{D_{d}}$
in future $\tau-c$ factory, and further has us get a good understanding of 
the $B-\bar B$ mixing phenomena. 

We would like to thank Professor R.J.Oakes for helpful discussions
and introducing Ref.\cite{4} to us. 
This work is in part supported 
by the National Science Foundation
of China.


    
\end{document}